\documentclass[aps,prc,12pt,amsmath,amssymb,nofootinbib,superscriptaddress,preprint]{revtex4-1}
\usepackage{selinput}
\usepackage[T1]{fontenc}
\usepackage{graphicx}
\usepackage{verbatim}
\usepackage[dvipsnames]{xcolor}
\usepackage{geometry}
\usepackage{units}
\usepackage[toc,page]{appendix}\usepackage{float}
\usepackage{subcaption}
\usepackage{lipsum}
\usepackage{hyperref}
\usepackage{caption}
\usepackage{bm}
\newcommand \beq{\begin{equation}}
\newcommand \eeq{\end{equation}}

\begin{document}
\title{A New State of Matter between the Hadronic Phase and the Quark-Gluon Plasma{\Large ?}}

\author{Yuki Fujimoto}
\affiliation{Department of Physics, University of California, Berkeley, CA 94720, USA}
\affiliation{RIKEN Center for Interdisciplinary Theoretical and Mathematical Sciences (iTHEMS), RIKEN, Wako 351-0198, Japan}

\author{Kenji Fukushima}
\affiliation{Department of Physics, The University of Tokyo, 7-3-1 Hongo, Bunkyo-ku, Tokyo 113-0033}

\author{Yoshimasa Hidaka}
\affiliation{Yukawa Institute for Theoretical Physics, Kyoto University, Kyoto 606-8502, Japan}
\affiliation{RIKEN Center for Interdisciplinary Theoretical and Mathematical Sciences (iTHEMS), RIKEN, Wako 351-0198, Japan}

\author{Larry McLerran}
\affiliation{Institute for Nuclear Theory, University of Washington, Box 351550, Seattle, WA 98195, USA}

\preprint{YITP-25-80, N3AS-25-009, RIKEN-iTHEMS-Report-25}

\newcommand{\YF}[1]{\textcolor{magenta}{#1}}
\newcommand{\KF}[1]{\textcolor{teal}{#1}}
\newcommand{\LM}[1]{\textcolor{Red}{#1}}
\newcommand{\YH}[1]{\textcolor{blue}{#1}}

\newcommand{\Tc}{T_{\mathrm{c}}}
\newcommand{\Td}{T_{\mathrm{d}}}
\newcommand{\Th}{T_{\mathrm{H}}}
\newcommand{\Nc}{N_{\mathrm{c}}}
\newcommand{\Nf}{N_{\mathrm{f}}}
\newcommand{\Nl}{N_{\mathrm{l}}}
\newcommand{\Tr}{\mathop{\mathrm{Tr}}}
\renewcommand{\Re}{\mathop{\mathrm{Re}}}
\newcommand{\MeV}{\;\text{MeV}}
\newcommand{\GeV}{\;\text{GeV}}
\newcommand{\muB}{\mu_\mathrm{B}}
\newcommand{\muBhat}{\hat{\mu}_\mathrm{B}}
\newcommand{\MB}{M_\mathrm{B}}
\newcommand{\llimit}{\nearrow}

\begin{abstract}
Lattice-QCD simulations and theoretical arguments hint at the existence of an intermediate phase of strongly interacting matter between a confined hadron gas and a deconfined Quark-Gluon Plasma (QGP).  We qualitatively and semi-quantitatively explore and differentiate the phase structures in the temperature window from the QCD pseudo-critical temperature $\Tc\simeq 160\MeV$ to the pure-gluonic deconfinement temperature $\Td\simeq 285\MeV$.  We propose a three-regime picture using a hadron resonance gas (HRG) description augmented with the exponential spectrum of strings, corresponding to highly excited mesons and glueballs, based on the analysis of a large number, $\Nc$, of colors.  We estimate the entropy density from our model to confirm that the lattice-QCD data are bracketed with three regimes, i.e., a hadron gas, a QGP, and a new phase for $\Tc \lesssim T \lesssim \Td$.  In this new phase that we name a Spaghetti of Quarks with Glueballs (SQGB), thermal degrees of freedom of quarks are deconfined, yet gluons remain confined in glueballs.  Since the Hagedorn temperature, $\Th\sim 285\MeV$, is universal in the meson and the glueball sectors, in the infinite $\Nc$ limit, the phase diagram in the plane of the baryon chemical potential and the temperature is reduced to one with the confined and deconfined phases and Quarkyonic Matter at high density.  At large but finite $\Nc$, an SQGB window may open between these phases.  We point out that the SQGB has interesting similarities with Quarkyonic Matter and that this matter in the large $\Nc$ limit is confined as measured by the interaction between heavy quarks, but behaves in other respects like a quasi-free gas of quarks.  As a result of the extrapolation to $\Nc=3$, we present a revised phase diagram with the SQGB phase bounded by thermal crossovers.  Finally, we give a quantitative analysis of chiral symmetry restoration in the SQGB phase.
\end{abstract}

\maketitle

\section{Introduction}

It is a longstanding and yet unsolved mystery how to characterize confinement and deconfinement of quarks and gluons at low and high temperatures in quantum chromodynamics (QCD).  One may think that the Polyakov loop in the fundamental representation is an approximate order parameter for deconfinement, and center symmetry is spontaneously broken at high enough temperature~\cite{Svetitsky:1985ye}.  However, the fundamental Polyakov loop can only control the excitation of quarks that belong to the fundamental representation of color.  The adjoint representation is not faithful, and gluons are insensitive to the center elements.  Therefore, the Polyakov loop in the adjoint representation takes a non-vanishing expectation value regardless of the realization of center symmetry.  In principle, the nature of the confinement-deconfinement transition may qualitatively be different for quarks and gluons, exhibiting distinct transition temperatures.  It is folklore that hadrons and glueballs as bound states of quarks and gluons, respectively, must melt simultaneously.

Our present work is inspired by a recent speculative scenario proposed by Cohen and Glozman~\cite{Cohen:2023hbq}.  We would stress that our interpretation is further pushed toward a nontrivial direction at the conceptual level along the lines of Quarkyonic Matter~\cite{McLerran:2007qj,Andronic:2009gj,Fujimoto:2023mzy}; see Ref.~\cite{Fukushima:2013rx} for a review including Quarkyonic Matter.  We will demonstrate that lattice-QCD data imply the existence of three regimes across thermal crossovers in which distinct but dual thermal degrees of freedom are matched.  Color deconfinement, in contrast, is physically realized as a consequence of the Debye screening.  From this novel point of view, we will revise the QCD phase diagram.

In the conventional arguments of phase classification, in the presence of dynamical quarks, center symmetry is explicitly broken, and the fundamental Polyakov loop is always finite.  There is no rigorous way to describe the confined phase at all.  The pseudo-critical temperature is often prescribed, but its exact value has no strict meaning.  As a matter of fact, there are a countless number of unsuccessful theoretical attempts to define the confinement order parameter for full QCD, including light quarks~\cite{Detar:1982wp,MEYERORTMANNS198431,Weiss:1987mp,Faber:1995up,Fukushima:2002bk,Ghanbarpour:2022oxt}.  So far, no handy measure to define finite-temperature confinement has been found.  The lack of an exact order parameter prevents us from a precise characterization of phase structures in a traditional way according to symmetry realization.

Nevertheless, suggestive observations from lattice-QCD simulations have encouraged theorists to conjecture more nontrivial structures between a hadronic phase and a deconfined Quark-Gluon Plasma (QGP).
It is a well-known fact that the pure gluonic theory with $\Nc=3$, where $\Nc$ stands for the number of colors, has a first-order phase transition at $T=\Td\simeq 285\MeV$~\cite{Borsanyi:2022xml, Giusti:2025fxu}.  Once dynamical quarks are implemented in QCD, the chiral phase transition is found around $T=\Tc\simeq 150$--$160\MeV$~\cite{Borsanyi:2013bia, HotQCD:2014kol}.  Moreover, the general tendency of locking between the chiral phase transition and deconfinement seems to persist in accessible regions of the QCD phase diagram.  However, it would be rather unnatural if the temperature window between $\Tc$ and $\Td$ is only an empty desert without any distinguishing characteristics.
We will enumerate several hints to disclose more structures for $\Tc \le T \le \Td$.

The first example of nontrivial behavior above $\Tc$ is found in the $T$ dependence of the trace anomaly.  As first pointed out in Ref.~[37] (footnote) of Ref.~\cite{Pisarski:2006hz}, if the dimensionless trace anomaly or the interaction measure is multiplied by $T^2$, this $T^2$-scaled trace anomaly is nearly constant in the temperature window of (1--4)$\Tc$ [or (1--4)$\Td$ in the pure gluonic theory].  This approximate scaling law implies some non-perturbative condensates with mass dimension 2 even in the deconfined phase.
Another example is the spectral peak of the heavy quarkonium, which remains even above $\Tc$~\cite{Asakawa:2003re,Burnier:2013nla,Petreczky:2021zmz,Lowdon:2022xcl,Bala:2023iqu}.  Some states of the bottomonium melt for $T>350\MeV$ or some could still survive at higher temperatures, while the Polyakov loop correlation shows flattening of the heavy-quark potential.
There are even more examples including heavy-ion collision phenomenology --- strongly-correlated QGP (sQGP) is a common phrase to refer to non-perturbative properties of matter above $\Tc$; see e.g.\ chapter 7 in Ref.~\cite{Gross:2022hyw} for a review of experimental data.  We point out that one possible interpretation for the sQGP is magnetically dominated and thus electrically correlated matter; see Fig.~1 of Ref.~\cite{Liao:2006ry} (see also Refs.~\cite{Liao:2008jg, Shuryak:2014zxa}).
From these theoretical, phenomenological, and even experimental discussions, it is evident that a single deconfinement order parameter is not adequate to identify the physical states for $T>\Tc$, and that we need to differentiate the properties of hot matter.
Along these lines, a partially confined/deconfined new phase has also been found in a model with the adjoint Polyakov loop~\cite{Myers:2007vc}, and a more general argument in large-$\Nc$ QCD has been proposed based on the lattice-QCD data~\cite{Hanada:2018zxn,Hanada:2020uvt,Hanada:2023krw,Hanada:2023rlk}.
To take account of the non-perturbative nature around $\Tc$, a center-symmetry projection method has been proposed~\cite{Vuorinen:2006nz}, which has invoked a notion of the semi-QGP~\cite{Hidaka:2008dr,Hidaka:2020vna} that is an alternative interpretation of the sQGP.

Recently, additional support for a new phase has been reported from the spectroscopy study in lattice QCD as emphasized by Cohen and Glozman~\cite{Cohen:2023hbq}; unexpected degeneracy among vector channels, $\bar{\psi}\tau^a \otimes \gamma_5\gamma^0\gamma^i\psi$,\, $\bar{\psi}\tau^a \otimes \gamma^0\gamma^i\psi$, and $\bar{\psi}\tau^a \otimes\gamma^i\psi$, has been observed in a particular temperature window of $\Tc \lesssim T \lesssim 3\Tc$~\cite{Rohrhofer:2019qwq, Rohrhofer:2019qal}.  This is not a parity doublet by $\gamma_5$, but the symmetry generators seem to involve $\gamma^0$, and hypothetical $SU(2)$ generated by $\{\gamma^0, -i\gamma_5\gamma^0, \gamma_5\}$ is called chiral spin symmetry~\cite{Glozman:2014mka,Glozman:2015qva,Glozman:2022lda}.
To justify such symmetry realization, Lorentz symmetry should be violated and only the temporal direction should become distinct.  One possible explanation for dynamically projecting the temporal direction out is the dominance of chromo-electric fluxes in this specific temperature window.  Indeed, such dominance of chromo-electric fluxes was seen around $T \simeq 2\Tc$ on the lattice within electrostatic QCD (EQCD), contrary to the perturbative ordering in which the electric screening is more efficient compared to the magnetic screening~\cite{Hart:2000ha}.  This picture has been emphasized more explicitly with the terminology of ``stringy fluid'' in Refs.~\cite{Rohrhofer:2019qwq, Rohrhofer:2019qal}.

The idea of the stringy fluid is naturally understood in the center-vortex picture of confinement.  One intuitive view of confinement-deconfinement change arises based on the percolation properties of such configurations with center fluxes~\cite{Engelhardt:1999wr}.  Interestingly, a recent lattice-QCD simulation has provided evidence for the second transition around $2\Tc$ up to where the vortex percolation persists~\cite{Mickley:2024vkm}.

These arguments coherently suggest a new phase regime for $T>\Tc$, but so far, the discussions have focused only on formal aspects.  This paper aims to explore the reality of the new phase regime between a hadron gas and QGP in terms of thermal degrees of freedom.  Going beyond the former discussions at zero baryon density, we will also generalize our considerations to the $\muB$-$T$ plane with $\muB$ the baryon chemical potential.  We will demonstrate a concrete model that is consistent with the HRG and is extendable to large $\Nc$ using the spectra of the open and closed strings.  We will find that such a very simple model provides us with a baseline to understand the lattice-QCD data and predicts a window of the new phase regime for large $\Nc$ as well as real QCD at $\Nc=3$.
This model of the intermediate phase turns out to be useful in understanding chiral symmetry restoration, and we will argue how the model can give a reasonable interpretation of the lattice-QCD data.

\section{Summary of Our Results}

We shall here summarize our results for the convenience of readers, and the subsequent sections will present the detailed justification and derivation of these results.

\subsection{$\Nc$-counting arguments}

The idea of an intermediate phase is inspired by the counting of $\Nc$ in thermodynamics.  At low temperatures, the energy density, the pressure, and the entropy density of a gas of color-confined hadrons are independent of $\Nc$, i.e., of $\mathcal{O}(\Nc^0)$.  A gas of quarks and a gas of gluons give thermodynamics of $\mathcal{O}(\Nc)$ and $\mathcal{O}(\Nc^2)$, respectively, at temperatures high enough to liberate color degrees of freedom.  Therefore, as the temperature continuously increases, matter should first go through a region where the energy density is of $\mathcal{O}(\Nc^0)$, and then should transit to a partially-deconfined region with the energy density of $\mathcal{O}(\Nc)$, and should eventually reach a fully-deconfined phase with gluons that lift the energy density to $\mathcal{O}(\Nc^2)$.  This is the most natural expectation unless a first-order phase transition discontinuously shifts the energy density at the critical temperature from $\mathcal{O}(\Nc^0)$ to $\mathcal{O}(\Nc^2)$.  We will see below that this latter case is possible in the large-$\Nc$ limit of QCD, while the phase diagram at finite $\Nc$ can accommodate an intermediate phase regime.

We can understand the dynamics of these phases by the nature of meson and glueball interactions.  Extensive thermodynamic properties for weakly interacting systems are characterized by the scattering amplitudes multiplied by the thermal phase-space density of particles involved in the scattering.  The strength of these interactions may lead us to the criteria of phases based on the large-$\Nc$ counting~\cite{Meyer:2009tq}.  For example, the amplitude for two-meson scatterings is of $\mathcal{O}(\Nc^{-1})$.  If we require this scattering amplitude times the meson density squared to be of $\mathcal{O}(\Nc)$, the meson density should be of $\mathcal{O}(\Nc)$.  This is the criterion that interactions are of equal strength in $\Nc$ to the kinetic energy of freely propagating mesons.  If the interactions are not strong enough to achieve this, the mesons may be regarded as weakly interacting.  In this case, we can characterize these mesonic contributions to extensive thermodynamic quantities by a non-interacting hadron resonance gas.  In contrast, the strength of glueball interactions is of $\mathcal{O}(\Nc^{-2})$, and glueballs are not important in thermodynamics until the glueball density becomes of $\mathcal{O}(\Nc^2)$.  We note that the interactions between mesons and glueballs are also of $\mathcal{O}(\Nc^{-2})$ and they are not important either until the densities of mesons and glueballs are saturated as $\mathcal{O}(\Nc)$ and $\mathcal{O}(\Nc^2)$, respectively, so that the amplitude times the densities can reach $\mathcal{O}(\Nc)$.

This argument means that there are three regions physically.  In one region in the low energy-density world, extensive thermodynamic quantities scale as $\mathcal{O}(\Nc^0)$, which is described by a Hagedorn gas of mesons and that of glueballs.  We will later consider the baryon contribution at finite density.  In the intermediate phase at higher temperatures where extensive quantities are of $\mathcal{O}(\Nc)$, the system may be thought of as a gas of quarks and a Hagedorn gas of glueballs.  In the region of the highest energy density, the system is fully deconfined to a gas of quarks and gluons.  We should note that the contribution of glueballs is suppressed in both low-density and intermediate phases because the glueball mass is as large as $M_\mathrm{GB} \simeq 1.65\GeV$ (see, e.g., Ref.~\cite{Athenodorou:2020ani} for the recent lattice-QCD calculation).  We will see that the transition to the highest-density phase with gluons occurs at the deconfinement temperature, $\Td \simeq 285\MeV$.  Only at this temperature, the density of states of glueballs increases such that glueballs melt, even though the Boltzmann factor of glueballs is still tiny.  Then, the contribution of gluons becomes relevant in thermodynamics.

\subsection{Thermal degrees of freedom and confinement}

In the present work, we have computed the contributions to thermodynamics in these various regions.  To model a gas of mesons and that of glueballs in such a way that is extendable to large $\Nc$, we employ the Hagedorn spectrum of open strings for mesons and that of closed strings for glueballs.  We provide a brief overview in Sec.~\ref{sec:string} to explain the derivation of the string formulae for the Hagedorn spectra.  It is clear from the derivation that the Hagedorn temperatures for the open and closed strings should be the same, which is given by
\begin{equation}
    \Th^2 = \frac{3}{2\pi} \sigma \,,
\end{equation}
where $\sigma$ is the string tension.  Using $\sqrt{\sigma} \simeq 440\MeV$, we have an estimate of $\Th \simeq 300\MeV$.  The string approximation is justified in the large-$\Nc$ limit of QCD, and there should appear finite-$\Nc$ corrections to the deconfinement temperature, i.e.,
$\Td/\sqrt{\sigma}\simeq 0.596(4)+0.453(30)/\Nc^2$~\cite{Lucini:2003zr}.  Therefore, 
$\Th/\Td\simeq 1.16 - 0.88/\Nc^2 \simeq 1.06$ for $\Nc=3$.  The direct fit in the lattice numerical simulation indicates a slightly smaller value of $\Th/\Td\simeq 1.024(3)$~\cite{Meyer:2009tq}.  This small deviation between $\Th$ and $\Td$ is associated with the first-order phase transition~\cite{Caselle:2015tza}.  In this work, we simplify the analysis and take the Hagedorn temperature close to $\Td$.  More specifically, we assume
\begin{equation}
    \Th \simeq \Td \simeq 285\MeV
\end{equation}
throughout this paper with the understanding that this value has uncertainties of $10\%$ at least for different $\Nc$ choices.
For $\Td$, we adopt the value obtained in the recent precision lattice study~\cite{Borsanyi:2022xml}, which is converted to the physical quantities using the scale parameter determined in Ref.~\cite{BMW:2012hcm}.

We are proposing a revised view of the phase diagram based on the argument of thermal degrees of freedom measured by the entropy density.  In Sec.~\ref{sec:entropy}, we shall derive the formula for the entropy density using the Hagedorn spectra.  The contribution from the low-lying states, namely, the light Nambu-Goldstone (NG) mesons such as pions and kaons should be handled separately from the sum over the Hagedorn spectra of heavy mesons and glueballs.  The non-NG modes are heavy compared to $\Th$, and thus the non-relativistic approximation works well to simplify the analytical treatment of the Hagedorn resonance gas.  In numerical comparisons, we will integrate the full relativistic expressions to evaluate the entropy density, while the non-relativistic simplification of the analytical expression is useful for us to deepen the qualitative understanding as discussed in Sec.~\ref{sec:entropy}.

A notable feature is found in the rising behavior of the entropy density in various regions.
The entropy density evaluated by the open-string spectrum corresponding to the meson resonance gas diverges at $T = \Th$.  In the strict limit of $\Nc\to\infty$, hence, the thermal degrees of freedom are saturated as $\mathcal{O}(\Nc)$ only at $T=\Th$.  For $\Nc=3$, however, at a much lower temperature, i.e., $\Tc \simeq 150$--$160\MeV$, the open-string (meson) contribution to the entropy density can be comparable to that from an ideal gas of quarks.
If we require chiral spin symmetry, chiral symmetry should be restored there, while the quark masses might still be close to the constituent masses rather than the bare ones.
We note that medium effects can give rise to a thermal mass not breaking chiral symmetry~\cite{Bellac:2011kqa}, so that quarks can acquire the thermal masses in the perturbative hot-QCD treatment.

Liberation of quark degrees of freedom in our proposed scenario is indeed consistent with the conventional picture of the QCD phase diagram. 
Beyond this point of $\Tc$, the system has thermal degrees of freedom of quarks and glueballs and the entropy density is dominated by contributions from an ideal gas of quarks and the closed-string (glueball) spectrum.  Interestingly, the entropy density of the closed-string (glueball) gas remains finite even in the limit of $T \llimit \Th$\footnote{The symbol $a\llimit b$ means that $a$ approaches $b$ from below.}, although susceptibilities involving higher-order derivatives may diverge.

\begin{figure}
    \includegraphics[width=0.7\linewidth]{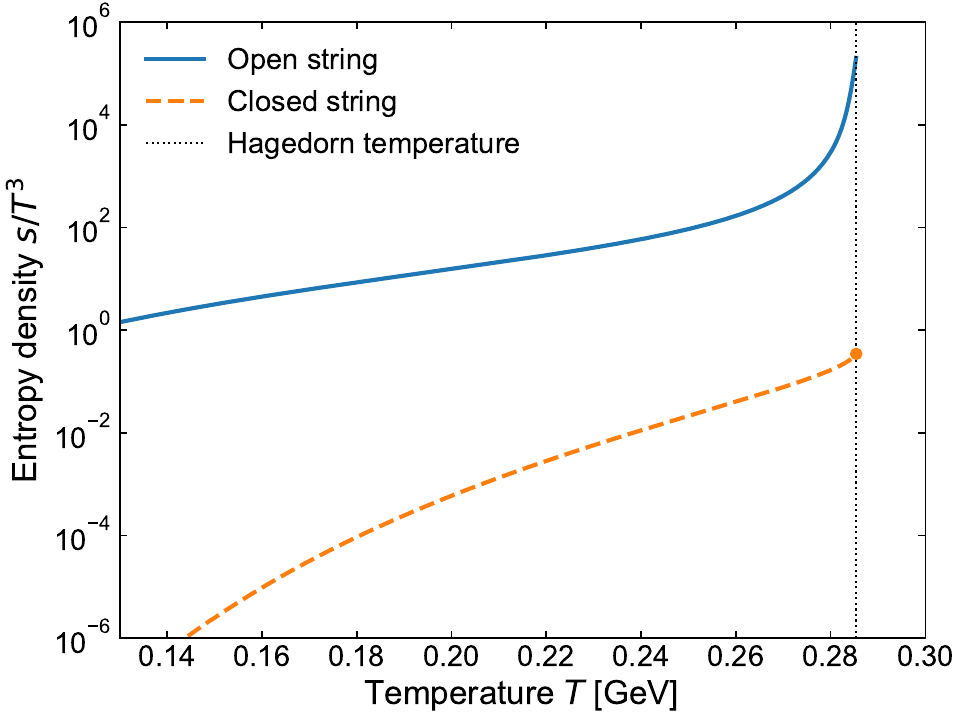}
    \caption{Comparison between the entropy densities from an open-string gas (drawn by the solid curve) and a closed-string gas (drawn by the dashed curve) up to the Hagedorn temperature of $\Th = 285\MeV$.}
    \label{fig:openclosed}
\end{figure}

In Fig.~\ref{fig:openclosed}, we plot the contribution of $s/T^3$ numerically evaluated from an open string gas and a closed string gas using the results in Secs.~\ref{sec:Hagedorn} and \ref{sec:model}.  It is evident that Fig.~\ref{fig:openclosed} explicitly shows the smallness of the closed-string (glueball) contribution.  This figure verifies the aforementioned descriptions about the rising behavior of the entropy density.

In summary, our picture from the low-$T$ to the high-$T$ regions is sketched as follows.
For a pure gluonic theory, there is a first-order phase transition for $\Nc\ge 3$~\cite{Svetitsky:1985ye}. 
In the closed-string phase where gluons are confined, the entropy density of closed strings or glueballs is of $\mathcal{O}(\Nc^0)$.  This means that the glueballs remain weakly coupled and give only a minor contribution to the entropy density.  Above the temperature $\Td \sim \Th$, the system must convert into a gas of gluons in the deconfined phase.  When dynamical quarks are included, this transition is presumably smoothed into a continuous crossover around the Hagedorn temperature for $\Nc = 3$. For the gluon gas just above $\Th$, we expect that the typical gluon mass would still be $\sim M_\mathrm{GB}/2$ due to the continuous nature of crossover, and again, the gluon contribution to the entropy density should be subdominant, so the effects of such a transition are rather small.  Here, above $\Th$, we can reasonably model the system in terms of an ideal gas of massive gluons and (chiral-symmetric) massive quarks.  We point out that the lattice simulation supports such modeling.  In fact, for a pure gluonic theory, the direct lattice calculation shows a good quantitative agreement with the entropy density from the closed-string gas model~\cite{Meyer:2009tq} (see also Ref.~\cite{Trotti:2022knd}).  Since the phase transition is of first order\footnote{As a trace of the first-order phase transition in the pure gluonic theory, for QCD with dynamical quarks, at $\Th$ there might be a divergence in the derivatives of $s/T^3$, viz., the specific heat $C_V$, or perhaps in some higher order derivative.  Currently, such divergence has not been seen evidently in the lattice-QCD calculation of $C_V$~\cite{HotQCD:2014kol}.}, the entropy density jumps at $\Td$ by
\begin{equation}
  \Delta s / T^3 \simeq 1.45\,.
\end{equation}
This is smaller than the value expected for a gas of massless gluons;  below the transition, $s/T^3\mid_\text{closed string} \, \simeq 0.3$ (see Fig.~3 in Ref.~\cite{Meyer:2009tq}) and above $\Td$ we can easily estimate $s/T^3\mid_\text{massless gluons}\, \simeq 7.02$ [see Eq.~\eqref{eq:s_gluon}] for $\Nc=3$.  Thus, gluon excitations should be suppressed.  For a gas of massive gluons with $m_\mathrm{gluon} \sim M_\mathrm{GB}/2 \simeq 0.8\GeV$, on the other hand, we get $s/T^3 \simeq 2.98$, which is still a somewhat larger value than suggested by the lattice data.  This mismatch indicates that the effective gluon mass at $\Td$ might even be larger than $0.8\GeV$.  We will nevertheless adopt the value of $m_\mathrm{gluon} = 0.8\GeV$ throughout this work.
Actually, in the quasi-particle model analysis, the effective gluon mass at the critical temperature is estimated as $m_\text{gluon}\sim (4$--$5)T$~\cite{Levai:1997yx}, which is consistent with our assumption.
The physical picture of the fate of gluons is amusing.  In the hadronic phase and in the intermediate phase that we will discuss below, the glueball contribution is very small.  Even at the critical temperature, $\Td\sim \Th$, the gluon contribution is relatively small, but it can eventually be non-negligible after deconfinement.  The deviation of the entropy density from the Stefan-Boltzmann limit is largely due to this depletion of gluons.

\begin{figure}
    \includegraphics[width=0.7\linewidth]{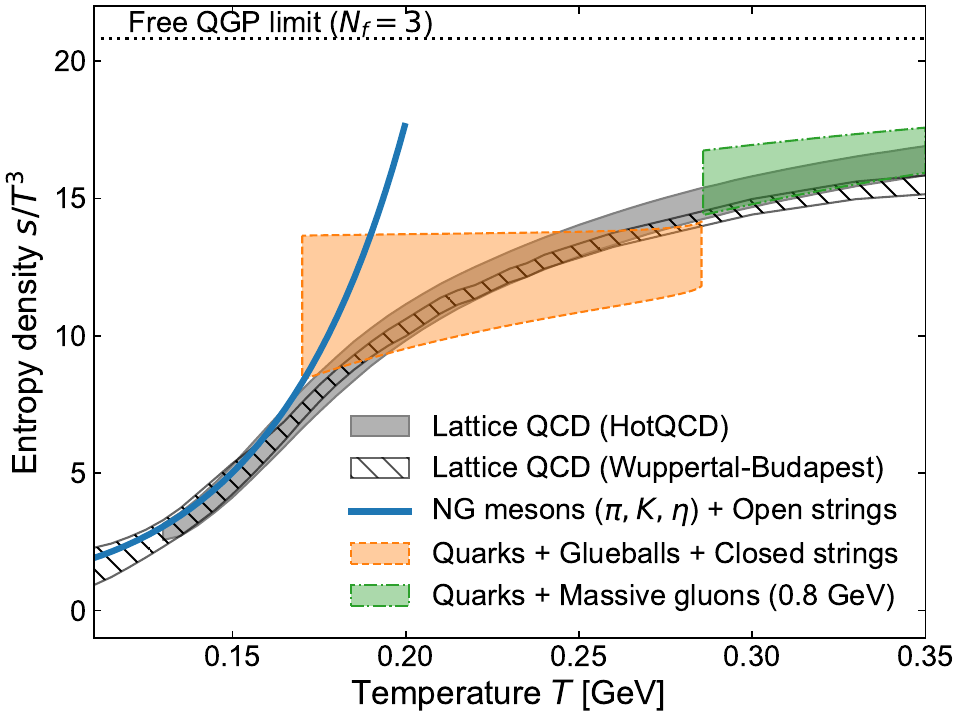}
    \caption{Lattice-QCD equation of state at high $T$ and zero density compared to the three-regime model.  In the low-$T$ hadronic regime, the sum of the low-lying mesons, the open-string (meson) gas multiplied by the flavor factor, and the closed-string (glueball) gas fits the lattice data well.  In the intermediate stringy regime, the thermodynamics is approximated by the contributions from glueballs and quarks with masses varied from the bare to the constituent values.  In the high-$T$ regime above $\Td$, the QGP with deconfined quarks and gluons with masses varied from zero to $M_\mathrm{GB}/2$ is presented.}
    \label{fig:3regimes}
\end{figure}

Figure~\ref{fig:3regimes} supports the validity of our three-regime model.  In the low-$T$ regime, as drawn by the solid curve, the sum of the low-lying mesons and the open-string gas (excited mesons) together with the negligibly small glueballs results in the entropy density in good agreement with the lattice-QCD data.  This is a little surprising, for the open-string spectrum grows with the Hagedorn temperature of $\Th = 285\MeV$, which is almost twice the crossover temperature, $\Tc$, from mesons into quarks.  The intermediate regime brackets the lattice data with the model computation dominated by an ideal gas of quarks with masses varied from the bare to the constituent values.  Even though the glueball spectrum follows the Hagedorn form, no divergence appears at $T=\Th$, as we already explained.  The high-temperature region above $\Th$ is also bracketed by a gas of quarks and gluons with masses varied from zero to half of the lightest glueball mass.  The model and computation leading to this figure will be detailed in Sec.~\ref{sec:model}.

\subsection{Deconfinement due to the screening effects}

So far, we have discussed how the thermal degrees of freedom are taken over by quarks and gluons in different regimes as $T$ increases.  However, it is important to state that the activation of quarks and gluons in thermodynamics does not necessarily mean their deconfinement.
The confinement properties, especially in the intermediate regime, are quite interesting.  Confinement is cut off due to the effect of Debye screening.  When gluons are tightly bound into glueballs, they should not make a thermal contribution to the Debye screening mass unless $T$ is comparable to the glueball mass.  This screening effect is generated only by quarks then, and in the large-$\Nc$ limit, the quark contribution should vanish.  We can understand this from the standard formula for the Debye screening mass:
\begin{equation}
    M^2_\mathrm{Debye} = g^2\biggl({1 \over 3} \Nc + {1 \over 6}\Nf\biggr)T^2\,,
    \label{eq:Debye}
\end{equation}
where $\Nf$ is the number of flavors.
The first term originates from gluon loops and the second from quark loops.  In the large-$\Nc$ limit, the 't~Hooft coupling, $g^2\Nc$, remains finite.  Thus, the first term in the above expression is finite, but if gluons are tightly bound into glueballs, this first term must be dropped.  Therefore, in the intermediate phase, the Debye screening mass squared is suppressed as $\mathcal{O}(\Nc^{-1})$, and quarks should still be confined, despite the fact that thermal degrees of freedom seem reasonably approximated by an ideal gas of massive quarks.
We also note that a similar $\mathcal{O}(\Nc^{-1})$ suppression takes place for the effect of finite quark chemical potential $\mu$.
Therefore, even if one goes to large $\mu$, the screening is unaffected by the existence of the quark medium.
This is essentially why $\Td$ stays constant at finite $\mu$ in the large-$\Nc$ limit, as will be discussed in the next section.

In this sense, the intermediate regime is quite similar to Quarkyonic Matter.  On the QCD phase diagram, Quarkyonic Matter is located at low temperature and finite baryon density.  Because the gluon degrees of freedom are not important there, confinement persists until a density scale, $\rho \sim \Nc^{3/2}\Lambda^3_\mathrm{QCD}$.  In Quarkyonic Matter, the energy density also scales as $\mathcal{O}(\Nc)$.  In both Quarkyonic Matter and the intermediate regime of our present interest, one should think of quarks as being joined together by color flux-tubes or color strings.  Yet, the density of stringy objects is large enough that these strings are intertwined.  One may say that the stringy configurations significantly increase the entropy density.  This can be translated into the contribution from the kinetic energy of quarks at the ends of strings, which dominates over the potential energy from stretching the strings~\cite{Bala:2021fkm}.  We will refer to this intermediate form of matter generically as a Spaghetti of Quarks with Glueballs or SQGB in short.  We will distinguish this matter from Quarkyonic Matter based on chiral properties.

In the high-$T$ phase where glueballs have evaporated into gluons, the system fully deconfines color charges, although due to the massive nature of glueballs and gluons near $\Td$, the full deconfinement may occur at somewhat larger temperatures than naively expected.  For $\Nc=3$, of course, these considerations are only qualitative,
but it remains true that the confinement scale is significantly larger in the intermediate phase than what would be expected from a gas of nearly massless gluons.  This presumably explains the non-perturbative nature of the lattice-QCD data at temperatures higher than $\Tc$, up to temperatures of order of the Hagedorn temperature, $\Th \sim 285\MeV$, or even higher.

\subsection{Chiral symmetry}

We briefly mention the chiral properties in the SQGB phase, which will be addressed in detail in Sec.~\ref{sec:chiral}.  The main content of that section is that the chiral-symmetry order parameter, $\langle\bar{\psi}\psi\rangle \sim \mathcal{O}(\Nc)$, receives finite-$T$ contributions that scale with the quark density multiplied by a factor of $\mathcal{O}(\Nc^0)$.  When the quark density is of $\mathcal{O}(\Nc)$, the total corrections reduce the expectation value of $\langle\bar{\psi}\psi\rangle$ and it is quite plausible that in the SQGB phase chiral symmetry would be restored.  According to the arguments of Refs.~\cite{Glozman:2022lda,Cohen:2023hbq,Cohen:2024ffx}, chiral symmetry could be restored in a way with emergent chiral spin symmetry.

\section{Revised Phase Diagrams}

\begin{figure}
    \includegraphics[width=0.8\textwidth]{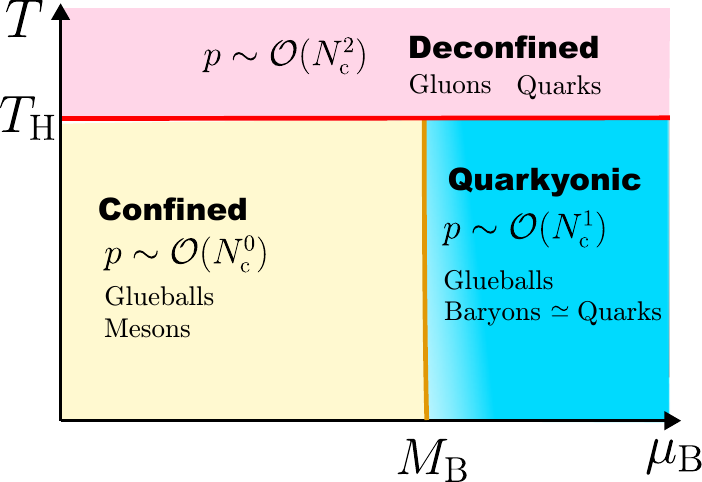}
    \caption{Quarkyonic phase diagram in the strict large-$\Nc$ limit.}
    \label{fig:diagram_largeN}
\end{figure}

How do we understand the relevance of the SQGB phase on the QCD phase diagram in the plane of the baryon chemical potential vs.\ the temperature?  First of all, we should clearly recognize that, in the strict large-$\Nc$ limit, the onset temperature of the SQGB will approach the QGP critical temperature.  This is because, as $\Nc$ diverges, the temperature of the string resonance gas will have to approach $\Th$ in order to saturate the thermodynamic degrees of freedom, i.e.,
\begin{equation}
  \lim_{\Nc\to\infty}~ \Td = \lim_{\Nc\to\infty}~ \Tc ~\to~ \Th\,.
\end{equation}
Another landmark is that the energy densities of Quarkyonic Matter and the SQGB phase share the same scaling of $\mathcal{O}(\Nc)$.  The onset of Quarkyonic Matter at $T\simeq 0$ lies at a baryon chemical potential, $\muB$, close to the baryon mass, $\MB$.  Therefore, in this strict large-$\Nc$ limit, there is no window of the SQGB on the phase diagram as depicted in Fig.~\ref{fig:diagram_largeN}.  It is still important to realize that in this infinite-$\Nc$ phase diagram, while the transition occurs at a single temperature, the entropy density changes between of $\mathcal{O}(\Nc^0)$ to $\mathcal{O}(\Nc^2)$, so that if one considers properties of matter as a function of the entropy density, they should drastically change.

\begin{figure}
    \includegraphics[width=0.8\textwidth]{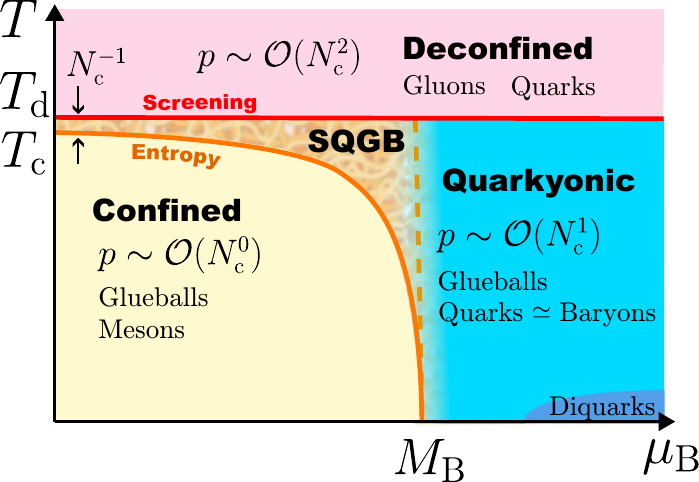}
    \caption{New phase diagram with stretched Quarkyonic Matter toward the temperature axis.  Along the temperature axis, a new window of the SQGB regime can open with a suppressed width $\sim \mathcal{O}(\Nc^{-1})$.}
    \label{fig:diagram_finiteN}
\end{figure}

For a large but finite value of $\Nc$, a small window of $\mathcal{O}(\Nc^{-1})$ opens between the deconfinement temperature $\Td\simeq \Th$ (characterized by the Debye screening) and the thermodynamic crossover temperature $\Tc$ (estimated by the entropy density).
Although thermal degrees of freedom in Quarkyonic Matter and the SQGB seem similar, the SQGB presumably has restored chiral symmetry, while chiral symmetry is weakly broken in Quarkyonic Matter.  Although naive arguments would also suggest chiral symmetry restoration in Quarkyonic Matter, interactions on the Fermi surface generate spatially inhomogeneous condensate that locally breaks chiral symmetry~\cite{Kojo:2009ha}.  The curvature of the SQGB transition temperature as a function of $\muB$ is small as indicated by the chemical potential dependence of the pseudo-critical temperature computed in the lattice-QCD simulation.  The curvature should be even smaller for the onset of the SQGB for a large value of $\Nc$.
The slope of the SQGB boundary in the high-density side can be estimated roughly by the condition that the baryon multiplicity $\sim e^{M_\mathrm{B}/\Th'}$ (where $\Th'$ for baryons may be different from $\Th$) is balanced by the thermal weight $\sim e^{(\muB-M_\mathrm{B})/T}$; see discussions in Sec.~3 of Ref.~\cite{Andronic:2009gj}.
Consequently, the expected phase structure should look like Fig.~\ref{fig:diagram_finiteN}.  For finite $\Nc$, generally, phases with diquark condensates may also occur at low-$T$ and high density.

\begin{figure}
    \includegraphics[width=0.92\textwidth]{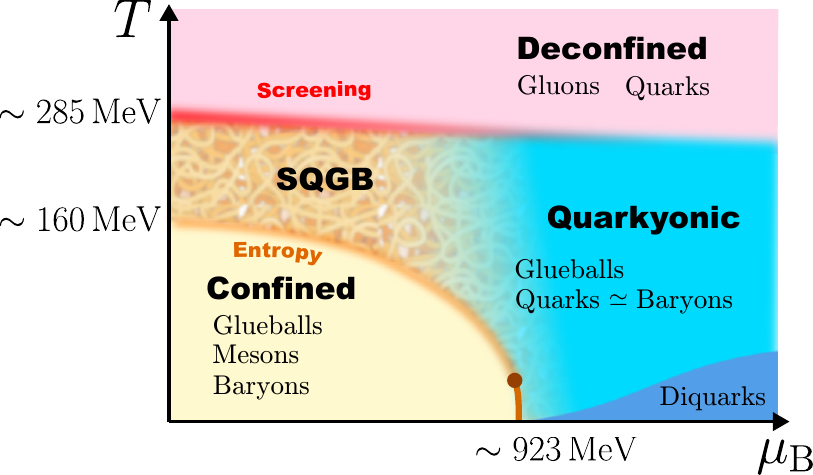}
    \caption{New and realistic phase diagram for $\Nc=3$ including a window of the SQGB regime.}
    \label{fig:diagram_N3}
\end{figure}

Now, we can make an extrapolation of $\Nc$ toward a realistic value of $\Nc=3$, in which all transitional changes should be smooth crossovers except for the liquid-gas transition of symmetric nuclear matter that is a first-order phase transition.  In Fig.~\ref{fig:diagram_N3}, we present our most plausible picture of the new phase diagram including a window of the SQGB regime.  In the conventional phase diagram, the counterpart of the onset of the SQGB is simply the pseudo-critical temperature $\Tc\sim 150$--$160\MeV$.  We claim that, even above $\Tc$ presumably up to $\Td\sim \Th \simeq 285\MeV$, the stringy phase continues with thermal degrees of freedom of quarks, even though quarks are still unscreened and thus confined.  We would emphasize that our picture does not contradict the preceding modeling efforts;  in terms of the HRG model, the repulsive interactions between mesons as considered in Ref.~\cite{Andronic:2012ut} are the ingredients for the model to confront lattice-QCD thermodynamics.  In our language, through interactions in such a model or the van~der~Waals parametrization of thermodynamics~\cite{Vovchenko:2016rkn,Fujimoto:2021dvn}, the system somehow gets to know the presence of quarks.

The temperature upper bound of the SQGB phase at a given $\muB$, viz., $\Td(\muB)$, is given by the screening argument based on the thermal screening mass expression~\eqref{eq:Debye}.
When $M_{\mathrm{Debye}}$ is comparable to the QCD scale or larger, the confining interaction is not long-ranged anymore due to screening, and thus the deconfinement sets in.
This is essentially the argument presented in Ref.~\cite{Bluhm:2024uhj}, in which Figs.~4 and 5 provide the estimate of $\Td(\muB)$.

\section{Hagedorn Spectrum}
\label{sec:Hagedorn}

In this section, we derive the expressions of the Hagedorn spectrum for the closed and open strings, and discuss the behavior of the entropy density for those Hagedorn spectra.  These are well-known results, but we provide them here to make this paper self-contained.

\subsection{Derivation of the closed and open string results for the Hagedorn spectra}
\label{sec:string}

We follow the arguments found in the standard textbooks; see Ref.~\cite{Green_Schwarz_Witten_2012}.
For the open string, the energy (mass) spectrum is given by
\begin{equation}
    \alpha' M^2 = N - 1\,,
\end{equation}
or in terms of the string tension, $\sigma$, instead of $\alpha'$, it is approximately written as
\begin{equation}
    \frac{M^2}{2\pi\sigma} \simeq N\,,
\end{equation}
where $N\gg 1$ is assumed.  Here, $N$ is the sum of the string modes, i.e.,
\begin{equation}
    N = \sum_{n=1}^\infty n \alpha_n^\dag \alpha_n\,,
\end{equation}
where $\alpha_n$ is the annihilation operator of right-moving (or left-moving) string modes with $n$-th quanta.  The physical states are prepared with the Fock states with transverse polarizations, so the number of states is the number of partitions of the total energy.  How to count this is given by the Hardy-Ramanujan formula (for more details, see Ref.~\cite{Green_Schwarz_Witten_2012}), and for the case with $b = D-2$ polarizations, the partition number is found to be
\begin{equation}
    p_b(N) \simeq \frac{1}{\sqrt{2}} \biggl( \frac{b}{24} \biggr)^{(b+1)/4} N^{-(b+3)/4} \exp\biggl( 2\pi\sqrt{\frac{Nb}{6}} \biggr)\,.
\end{equation}
For our problem of the four-dimensional theory, we use $b=2$ (though in the string theory textbook, $D=26$ is the standard choice to avoid anomalies without supersymmetry), and so we are interested in
\begin{equation}
    p_2(N) \simeq \frac{1}{2^2\cdot 3^{3/4}} N^{-5/4} e^{2\pi \sqrt{N/3}}\,.
\end{equation}
In the case of the \textit{closed} strings, corresponding to the glueballs in our problem, one must be careful of the energy spectrum that is given by
\begin{equation}
    \frac{M^2}{2\pi\sigma} \simeq 2 (N + \tilde{N})\,,
\end{equation}
and in this case, $N$ contains both left-moving and right-moving string vibrations, i.e.,
\begin{equation}
    N + \tilde{N} = \sum_{n=1}^\infty (n\alpha_n^\dag \alpha_n + n\tilde{\alpha}_n^\dag \tilde{\alpha}_n )\,.
\end{equation}
For the strings with zero momentum, the first term and the second term should be the same, that is, $N=\tilde{N}$.  This means that the physical states are given by the tensor product of the states with $\alpha_n^\dag$ and the states with $\tilde{\alpha}_n^\dag$.  So, the number of states is given by $[p_n(N)]^2$, that is,
\begin{equation}
    p_{\text{closed}}(N) = \frac{1}{2^4 \cdot 3^{3/2}} N^{-5/2} e^{4\pi\sqrt{N/3}}\,.
\end{equation}
Now, we use the following relations,
\begin{equation}
    M^2 = 8\pi\sigma N\,,\qquad
    M dM = 4\pi \sigma dN\,,
    \label{eq:dM}
\end{equation}
and $\rho_{\text{closed}}(M)dM = p_{\text{closed}}(N) dN$, the spectrum as a function of the mass, $M$, is finally given by
\begin{equation}
    \rho_{\text{closed}}(M) = \frac{(2\pi\sigma)^{3/2}}{3^{3/2}} M^{-4} e^{M \sqrt{2\pi/3\sigma}}\,.
\end{equation}
From this expression, we can identify the Hagedorn temperature for the closed string as
\begin{equation}
    T_{\text{closed}} = \sqrt{\frac{3\sigma}{2\pi}}\,.
\end{equation}
For convenience, we can also eliminate $\sigma$ using $T_{\text{closed}}$, and after all, we arrive at the formula used by Meyer~\cite{Meyer:2009tq}, that is,
\begin{equation}
    \rho_{\text{closed}}(M) = \frac{(2\pi)^3}{27 T_{\text{closed}}} \biggl(\frac{T_{\text{closed}}}{M}\biggr)^4 e^{M/T_{\text{closed}}}\,.
    \label{eq:rhoclosed}
\end{equation}

Obviously, the closed-string calculation is more non-trivial than the open-string case, and for the open string, we do not have to calculate $[p_2(N)]^2$ but simply $p_2(N)$ gives the partition number.  We should be just careful of the different relation between $M$ and $N$ as given at the beginning of this subsection.  It is a straightforward exercise to repeat the same procedure to find
\begin{equation}
    \rho_{\text{open}}(M) = \frac{\sqrt{2\pi}}{6 T_{\text{open}}} \biggl(\frac{T_{\text{open}}}{M}\biggr)^{3/2} e^{M/T_{\text{open}}}\,.
    \label{eq:rhoopen}
\end{equation}
Here, interestingly, the Hagedorn temperature is the same, i.e., $T_{\text{open}}=T_{\text{closed}}$, which is denoted by $\Th$ in this work.
We note that the result for the open string has been discussed in Ref.~\cite{Huang:1970iq}.\footnote{In Ref.~\cite{Huang:1970iq}, the authors apparently had a different exponent for the $M$ in the prefactor of the above expression as they did not take into account the relations corresponding to Eq.~\eqref{eq:dM}.}

\subsection{Entropy estimates from the Hagedorn model}
\label{sec:entropy}

The entropy density $s$ of the ideal gas with a given mass $M$ at $\mu = 0$ is
\begin{equation}
    s(M) = \frac{1}{6\pi^2 T} \int_0^\infty dk \, \frac{4 k^4 + 3 M^2 k^2}{\sqrt{k^2 + M^2}} \left[\exp\left(\frac{\sqrt{k^2 + M^2}}{T} \right) + \eta \right]^{-1}\,,
\end{equation}
where $\eta$ depends on the statistics: $\eta=1$ for the Fermi-Dirac and $\eta = -1$ for the Bose-Einstein statistics, respectively.

We can expand the entropy density as an infinite sum of the modified Bessel functions of the second kind, $K_\nu$, that is
\begin{equation}
    s(M) = \frac{M^3}{2\pi^2} \sum_{n=1}^{\infty} \frac{(-\eta)^{n+1}}{n} K_3\left(\frac{nM}{T}\right)\,.
    \label{eq:sBessel}
\end{equation}
It should be noted that taking only the $n=1$ term in the above expression corresponds to the Boltzmann approximation.

The entropy density for the open or closed string is thus given by
\begin{equation}
    s_\text{o/c} = \int_{M_0}^\infty dM\, d_\sigma (M) \rho_{\text{open/closed}}(M) s(M)\,,
    \label{eq:sstring}
\end{equation}
where $M_0$ is the lower bound of the mass and $d_\sigma(M)$ is the degeneracy of the string polarization at a given $M$.
The mass spectra for the open and closed string, $\rho_{\text{open/closed}}(M)$, can be found in Eq.~\eqref{eq:rhoopen} and Eq.~\eqref{eq:rhoclosed}, respectively.  We put one-letter subscripts, o and c, for notational brevity, to indicate the open and closed strings, respectively.

Let us make a non-relativistic approximation to compute $s_\text{o/c}$ in the zero baryon number density limit.  In this limit, Eq.~\eqref{eq:sBessel} takes a simple expression as
\begin{equation}
   s(M) ={1 \over {(2\pi)^{3/2}}} T^3 (M/T)^{5/2}e^{-M/T}\,.
\end{equation}
Then, we can derive the integral forms of the entropy densities:
\begin{equation}
   s_\text{o/c} = d_\text{o/c}\kappa_\text{o/c} T^3 \left({\Th \over T}\right)^{5/2} \; \int_{M_0} ~dM~ {1 \over \Th} \left({\Th \over M}\right)^{(a_\text{o/c} -5/2)} e^{M/\Th-M/T}\,,
\end{equation}
where for the open string we get,
\begin{equation}
    \kappa_\text{o} = {\sqrt{2\pi} \over 6}\,,
    \qquad
    a_\text{o} = 3/2\,,
\end{equation}
and for the closed string we get,
\begin{equation}
    \kappa_\text{c} = {{(2\pi)^3 \over {27}}}\,,
    \qquad
    a_\text{c} = 4\,.
\end{equation}
The lower mass $M_0$ is determined from the physical conditions;  $M_0$ for the open string is the lowest mass of the non-NG mesons and and $M_0$ for the closed string is the lightest mass of the non-low-lying glueballs.  The degeneracy of the meson states has a factor of $4$ for spin and a factor of $\Nf^2$ for flavors, which can be interpreted as the degeneracy of $q \bar{q}$ at the ends of the string.  In later calculations, when we compute the contribution to the entropy density, we will separately treat the non-strange and strange mesons, and we will take a flavor degeneracy of 4 for mesons made of up and down quarks, 4 for mesons with strangeness $\pm 1$, and 1 for mesons composed of an $s \bar{s} $ pair.

For $s_\text{o}$, we can perform the integration over mass and find the analytical result:
\begin{equation}
    s_\text{o} = d_\text{o}\kappa_\text{o} T^3 \left( {T_\mathrm{H} \over T} \right)^{3/2} \left( {{M_0(\Th-T) + \Th T}} \over {(\Th-T)^2}  \right) \exp\left(-{{M_0(\Th-T)}\over {\Th T}}\right)\,.
\end{equation}
This explicitly shows the Hagedorn singularity and the exponential suppression of the amplitude unless $T \sim \Th$.
In the limit of $T\llimit \Th$, we see that $s_\text{o}$ diverges.

The expression for the closed string is difficult to bring into a simple analytic form.  We note, however, that it is exponentially suppressed by the glueball mass, unless $T$ is close to $\Th$.  We encounter non-analytic behavior of the entropy density as we approach the Hagedorn temperature, and the first derivative of the entropy density with respect to $T$ diverges like $(\Th-T)^{-1/2}$ as $T \llimit \Th$.  In this limit of $T \llimit \Th$, interestingly, we see that $s_\text{c}$ remains finite:
\begin{equation}
 s_\text{c}\mid_{T = \Th}  \;=\; 2 d_\text{c}\kappa_\text{c} \Th^3 \sqrt{\Th/M_0}\,.
\end{equation}
We note in passing that such a qualitative difference in the behavior of the entropy density in the limit of $T \llimit \Th$, which depends on the exponent of the prefactor in the Hagedorn spectrum, was pointed out by Cabibbo and Parisi in their general analysis~\cite{Cabibbo:1975ig}.

We can also trace such an argument from Eq.~\eqref{eq:sBessel} by resorting to the relation:
\begin{align}
    \lim_{M \to \infty} (M/\Th)^a\, e^{M/\Th} K_3(M/\Th)
    =
    \begin{cases}
        0 & (a < 1/2)\,,\\
        \sqrt{\pi/2} & (a = 1/2)\,,\\
        \infty & (a > 1/2)\,.
    \end{cases}
\end{align}
The open string and closed string cases correspond to $a= 3/2$ and $a = -1$, respectively.
The former diverges, while the latter leads to the convergent integral over $M$ up to infinity.

\section{Supporting Argument Toward the SQGB Phase}
\label{sec:model}

We give a supporting argument toward the three-regime scenario \cite{Cohen:2023hbq} (the early incarnation can be found in Ref.~\cite{Stoecker:2015zea}) by showing a semi-quantitative comparison between the entropy density from the lattice QCD~\cite{Borsanyi:2013bia, HotQCD:2014kol} and from the Hagedorn spectrum.
We demonstrate that the Hagedorn temperature of $\Th \sim 285\MeV$ is inferred, and the temperature range slightly below this Hagedorn temperature is consistent with the system comprising free quarks with glueballs.
Also, we point out that the low-temperature behavior of the entropy density is well captured by the open string spectrum with the low-lying NG mesons singled out.

\subsection{Details of models}
Here, we describe the details of each contribution in the model: NG mesons, open strings, glueballs, closed strings, and free quarks.

\paragraph*{Nambu-Goldstone mesons:}
We single out the low-lying pseudo-Nambu-Goldstone (NG) mesons below $M=0.67\,\text{GeV}$, which are $\pi$, $K$, and $\eta$.
We only include the mesonic contribution below $M=0.67\,\text{GeV}$ in the HRG spectrum taken from Particle Data Group~\cite{ParticleDataGroup:2024cfk}.

\paragraph*{Open strings:}
For the open string Hagedorn spectrum~\eqref{eq:sstring}, we take $M_0 = 0.67\,\text{GeV}$ for the lower bound of the mass spectrum.
This value is the mass of the first non-NG meson, $K_0^\ast(700)$, in the HRG spectrum, and also corresponds to twice the constituent quark mass of light flavors ($u$ and $d$).
In Eq.~\eqref{eq:sstring}, we take the string polarization degeneracy factor $d_\sigma$ as follows depending on the different mass range:
\begin{align}
\label{eq:dsigma}
    d_\sigma(M) =
    \begin{cases}
        16 & (2m_l \le M < m_l + m_s)\,,\\
        32 & (m_l + m_s \le M < 2m_s)\,,\\
        36 & (2 m_s \leq M  < \infty)\,,
    \end{cases}
\end{align}
where $m_l \simeq 0.33\GeV$ and $m_s \simeq 0.49\GeV$ are the constituent mass for light quarks and strange quarks, respectively.
These degeneracy factor arises from quarks attached to the end of an open string; quarks attached to the string have $(2\Nf)^2$ degrees of freedom, where $2$ is for the spin, and $\Nf$ is for the number of quark flavors.
Below a single strangeness threshold mass value $m_l + m_s$, quarks attached to a string are both $\Nf=2$, which leads to $d_\sigma = 16$.
Likewise, when $m_l + m_s \leq M < 2m_s$, one of the quarks is treated as $\Nf=3$, and when $M \geq 2m_s$, both quarks are $\Nf = 3$.

\paragraph*{Glueballs:}
For the glueball and closed-string sectors, we follow the treatment in Ref.~\cite{Meyer:2009tq}.
We take the low-lying glueball spectrum from Ref.~\cite{Athenodorou:2020ani} below the two-particle threshold $2M_{\mathrm{GB}}$, where $M_{\mathrm{GB}}\simeq 1.65\,\text{GeV}$ is the mass of the lightest scalar glueball.
Above $2M_{\mathrm{GB}}$, we add the closed string Hagedorn spectrum contribution.

\paragraph*{Closed strings:}
In the expression for the closed string Hagedorn spectrum~\eqref{eq:sstring}, we take for the lower bound of the mass spectrum $M_0 = 2 M_{\mathrm{GB}}$.
For the string polarization degeneracy, we take the value $d_\sigma = 1$.
This is because all the degeneracies in the Hagedorn spectrum are incorporated in the string excitation.

\paragraph*{Free quarks:}
In Fig.~\ref{fig:3regimes}, we bracket the uncertainty of the quark mass.
For light flavors, we vary the mass between their bare quark masses $m_l = 0$ and constituent quark masses $m_l \simeq 0.33\GeV$.
For strange quark ($s$), we vary the mass between its bare mass $m_{s} \simeq 0.10\GeV$ and constituent mass $m_{s} \simeq 0.49\GeV$.
Particularly, the entropy density for massless quarks arises as
\begin{equation}
    \frac{s}{T^3} = \frac74 \nu_{q} \frac{2\pi^2}{45}\,,
\end{equation}
where the number of light quark degrees of freedom is $\nu_q = 2 \Nc N_l$, $N_l$ being the number of light flavors.
The RHS of the equation above amounts to $s/T^3 \simeq 4.6N_l$.

\paragraph*{Free gluons:}
In Fig.~\ref{fig:3regimes}, we assume the thermal mass for gluons, which we take as $0.8\GeV$.
Particularly, the entropy density for massless gluons is given by
\begin{equation}
    \frac{s}{T^3} = \nu_{g} \frac{2\pi^2}{45}\,,
    \label{eq:s_gluon}
\end{equation}
where the number of gluonic degrees of freedom is $\nu_g = 2(\Nc^2 - 1)$.
In the massless limit, the gluonic contribution to the entropy reaches $s/T^3 \simeq 7.0$.

\subsection{Comparison of the entropy densities}

In Fig.~\ref{fig:openclosed}, we showed the behavior of the entropy density for the open and closed strings.
As we have described above, the open string has a larger contribution compared to the closed string.
This is owing to the difference in the exponent of $M$ in the prefactor of the Hagedorn spectrum $\rho (M)$.
Also, such a difference will lead to the difference in the limit of $T \llimit \Th$ as already explained in Sec.~\ref{sec:entropy}: $s/T^3$ remains finite for closed strings, while it diverges for open strings.
The limiting value at $T\llimit \Th$ is shown by the circle marker in Fig.~\ref{fig:openclosed}.
Roughly speaking, degrees of freedom in the entropy density are proportional to the multiplicities of each contribution.
From Fig.~\ref{fig:openclosed}, we see that the multiplicity of glueballs (closed strings) is $10^{-3}$ times that of mesons (open strings).

\begin{figure}
    \centering
    \includegraphics[width=0.7\linewidth]{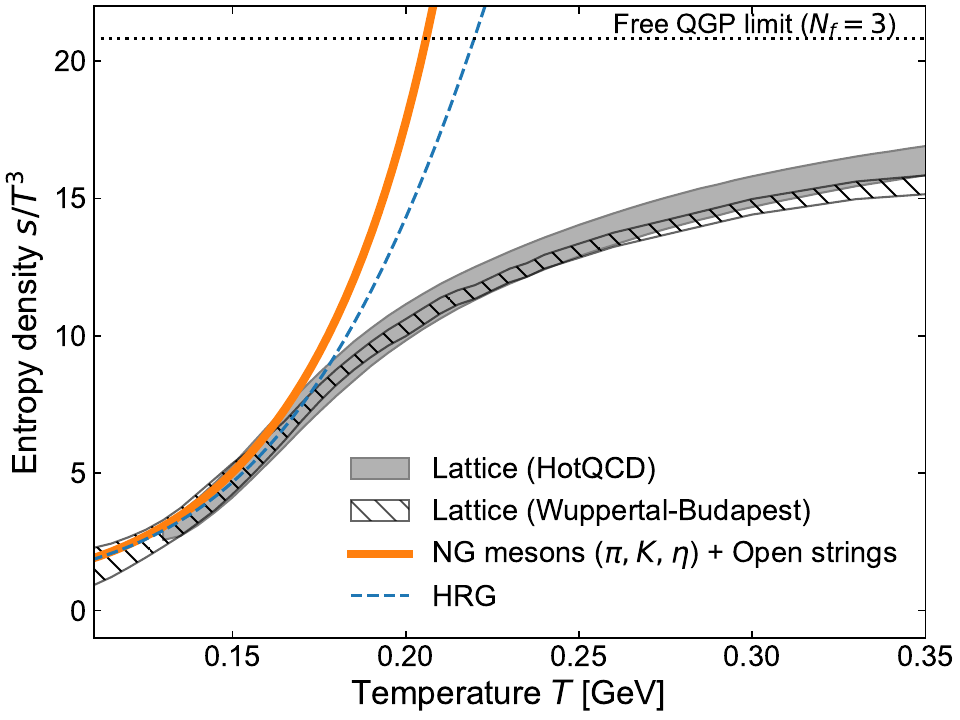}
    \caption{Comparison between the HRG model and the Hagedorn spectrum of the open strings.}
    \label{fig:open}
\end{figure}

In Fig.~\ref{fig:open}, we compare the HRG model with the entropy derived from the Hagedorn spectrum of the open string\footnote{We note that here the contributions from baryons may be missing.  The inclusion of the baryons will slightly increase the entropy density at a given $T$, and eventually leads to a modest increment in the Hagedorn temperature~\cite{Lo:2015cca}.  We note that the open string can describe some baryonic excitations on the Regge trajectories~\cite{Selem:2006nd}.}.
We observe that without any fine-tuning procedure, the open-string Hagedorn spectrum with the Hagedorn temperature of $\Th \simeq 285\MeV$ matches well with the lattice-QCD data and the HRG equation of state.
This observation is consistent with that there is only one Hagedorn temperature $\Th$ in QCD, common for mesons and gluons.
This may be that the experimentally observed hadron spectrum fitted with the Hagedorn exponential form, $e^{M/\Th}$, has the value $\Th = 285\MeV$.
For instance, the authors of Ref.~\cite{Broniowski:2000bj} indeed obtained the Hagedorn temperature $\Th \simeq 197\MeV$ for the meson spectrum (see also Refs.~\cite{Broniowski:2004yh, Lo:2015cca} for the updated estimate).
However, there is an ambiguity in the prefactor, which also depends on a power of $M$.
With an open string spectrum, we fix such an $M$-dependent prefactor independent of the specific model.
This direct application of the string formula suggests that QCD in the confined regime exhibits the stringy dynamics~\cite{Cohen:2009wq}.

\begin{figure}
    \centering
    \includegraphics[width=0.7\linewidth]{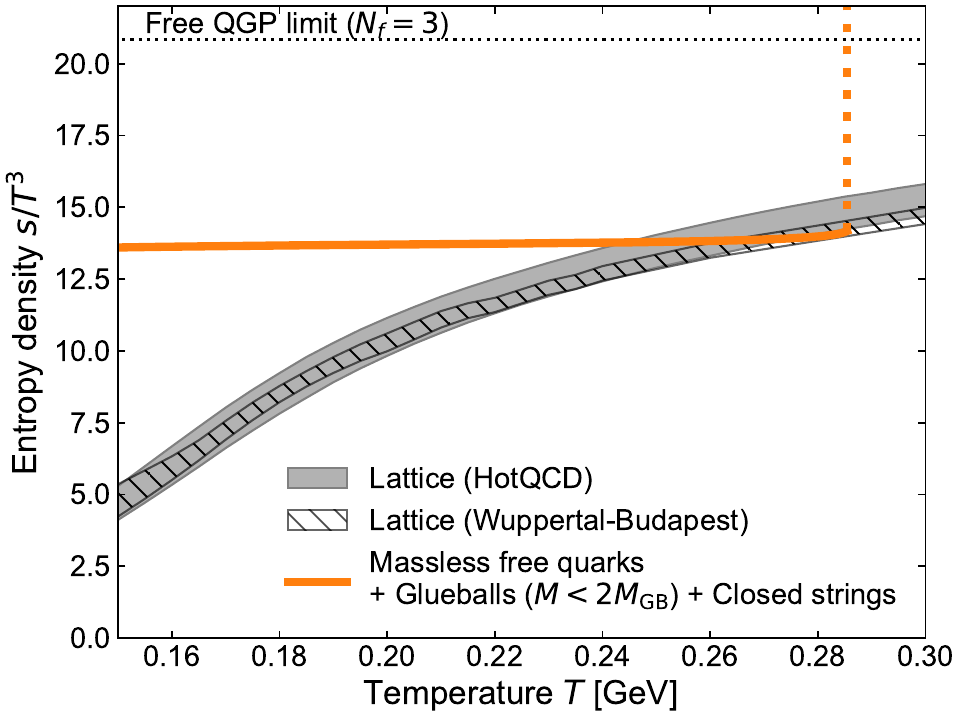}
    \caption{Comparison of the entropy density between the lattice-QCD data and the gas of massless free quarks and light glueballs as well as heavy glueballs that obey the closed-string spectrum.}
    \label{fig:closed}
\end{figure}

In Fig.~\ref{fig:closed}, we show the comparison between the entropy density of massless free quarks and glueballs, and lattice QCD\@.
Again, without any fine-tuning procedure, we observe these values are consistent with each other within uncertainty in a narrow window of temperature $\sim240$--$280\MeV$.
This observation supports the view that the SQGB phase can be described aptly by chirally symmetric massless quarks and glueballs.
As one increases the temperature, the interaction among quarks gets weaker, as is also implied from that the weak-coupling calculation in thermal QCD already works well around $\sim 2 \Tc$~\cite{Zhai:1995ac, Blaizot:2000fc, Kajantie:2002wa, Blaizot:2003iq, Laine:2006cp, Haque:2014rua}.

\subsection{Validity range of the temperature for the SQGB phase}

\begin{figure}
    \centering
    \includegraphics[width=0.7\linewidth]{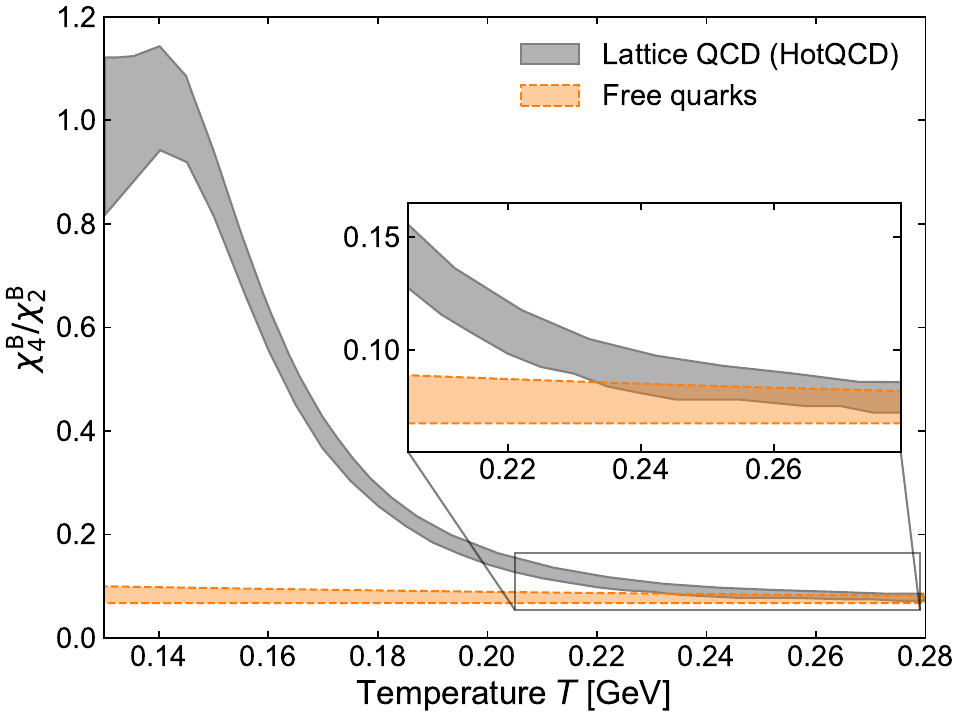}
    \caption{Comparison of the ratio of the fourth and second order cumulants of net-baryon number fluctuations ($\chi_4^{\mathrm{B}} / \chi_2^{\mathrm{B}}$) between the lattice-QCD data and the free quark gas value.}
    \label{fig:cumulant}
\end{figure}

Although we have bracketed the uncertainty for the SQGB phase (middle region) in Fig.~\ref{fig:3regimes}, there may be a limitation in the temperature window in which the description of the SQGB phase by free quarks and glueballs is apt.
As we have shown above, such a temperature range may reach down to $T\sim 240\MeV$, and the upper bound is at $T \sim \Td$.
As shown in Fig.~\ref{fig:cumulant}, this is in fact consistent with another lattice-QCD data on the behavior of the ratio of fourth and second order cumulants of net-baryon number fluctuations ($\chi_4^\mathrm{B}/\chi_2^\mathrm{B}$), where the susceptibility $\chi_i^{\mathrm{B}}$ is the $i$-th derivative of the pressure with respect to the dimensionless chemical potential $\muBhat = \muB / T$:
\begin{equation}
    \chi_i^\mathrm{B} = \left. \frac{\partial^i P(T, \muBhat)/ T^4}{\partial \muBhat^i} \right|_{\muBhat = 0}\,.
\end{equation}
The lattice-QCD data are taken from Fig.~3 in Ref.~\cite{Bazavov:2017dus} (see also Ref.~\cite{Borsanyi:2018grb}).
The band in the free quark value in Fig.~\ref{fig:cumulant} accounts for uncertainty in quark masses varied between the bare and the constituent values.
It should be noted that the cumulant ratio is sensitive to the quark content of thermal excitation and does not carry information about confining interactions.  Moreover, this measure is blind to the gluonic contribution.
Above $T\simeq 220\MeV$, the free quark description matches with the lattice-QCD data.
Therefore, the free quark description may only work at $T \gtrsim 220\MeV$.
Below this temperature, the glueball contribution is expected to remain valid, but the massless free quark description may be inadequate because the interaction between quarks cannot be neglected.

There are many examples where almost free-quark descriptions work, although the quarks remain confined.  An example is the ratio of hadron production to muon production in $e^+e^-$ annihilation.  Even at low energies, where the production cross section for hadrons is described by resonances, smearing the cross section over an energy scale of the order of the confinement scale gives a result in accordance with the quark model.  This is a consequence of quark-hadron duality.

\section{Chiral Restoration}
\label{sec:chiral}

Since the SQGB phase starts at $T\gtrsim\Tc$ and chiral symmetry is believed to be restored there, we should address the consistency of our scenario and build a model that allows for chiral symmetry restoration.

In general, even in the case of a hadron gas~\cite{Biswas:2022vat}, the thermal correction to the chiral condensate is to be evaluated by the differentiation of the pressure with respect to the quark mass $m_q$:
\begin{equation}
    \langle \bar{\psi}\psi \rangle_T =\langle \bar{\psi}\psi \rangle_0 -  \frac{\partial p}{\partial m_q}\,,
    \label{eq:chiralm}
\end{equation}
where $\langle \bar{\psi}\psi \rangle_0$ represents the vacuum part of the chiral condensate.
The pressure $p(M)$ of an ideal meson gas with mass $M$ at zero chemical potential is
\begin{equation}
    p(M) = - T \int \frac{d^3 k}{(2\pi)^3} \ln \left[1 -\exp\left({-\sqrt{k^2 + M^2}/T}\right)\right],
\end{equation}
whose  derivative with respect to $m_q$ can be calculated as
\begin{align}
    \frac{\partial p(M)}{\partial m_q}
    &= -\sigma_M\,\frac{M}{m_q}\frac{1}{2\pi^2}\int_0^\infty dk \frac{k^2 }{\sqrt{k^2+M^2}}
    \frac{1}{e^{\sqrt{k^2 + M^2}/T}-1} \notag\\
    &= -\sigma_M\,\frac{M}{m_q}\frac{TM}{2 \pi ^2}\sum_{n=1}^\infty\frac{1}{n } K_1\left(\frac{n M}{T}\right)\,.
    \label{eq:delP/delmq}
\end{align}
Here, $\sigma_M$ is a quantity called the sigma term, i.e., $\sigma_M= m_q(\partial M/\partial m_q)$.
For light NG mesons, using the Gell-Mann–Oakes–Renner relation~\cite{Gell-Mann:1968hlm},  
$m_\pi^2 f_\pi^2=-m_q\langle\bar{\psi}\psi\rangle_0$, we obtain:
\begin{equation}
    \sigma_{\pi} \frac{m_\pi}{m_q} = \frac{1}{2}\frac{\partial m_\pi^2}{\partial m_q}
    = -\frac{\langle\bar{\psi}\psi\rangle_0}{2f_\pi^2}\,,
\end{equation}
and thus, $\sigma_{\pi} =m_\pi/2$.
This NG-meson contribution leads to~\cite{Gasser:1987ah,Gerber:1988tt}
\begin{equation}
\label{eq:chiralT}
   \frac{\langle\bar{\psi}\psi\rangle_T}{\langle\bar{\psi}\psi\rangle_0}
   \simeq 1-\frac{1}{8}\frac{T^2}{f_\pi^2}\,,
\end{equation}
where we took the chiral limit ($m_q\to 0$) and assumed that there are three pions as the NG mesons.
Since $f_\pi^2\sim \Nc$ in the large-$\Nc$ counting, one may think that Eq.~\eqref{eq:chiralT} implies the critical temperature for vanishing $\langle\bar{\psi}\psi\rangle_T$ is $T_\mathrm{chiral}\sim f_\pi\sim \sqrt{\Nc}$ which diverges for $\Nc\to\infty$.
However, this is not the case in QCD with the Hagedorn spectrum~\cite{Biswas:2022vat}.
The heavier hadrons all contribute to the chiral condensate, which reduces $T_\mathrm{chiral}$.

Although the sigma term has been evaluated for several hadrons~\cite{Copeland:2021qni}, there is no generic formula of the sigma term for hadron resonances.  Here, let us make a simple assumption:
\begin{equation}
   \sigma_M \approx n_l \bar{\sigma}\,,
\end{equation}
where $\bar{\sigma}$ is the effective sigma term per quark,
and $n_l$ represents the number of light quarks,
e.g., $n_l=2$ for mesons made of up and down quarks, $n_l=1$ for mesons with strangeness $\pm 1$, and $n_l=0$ for mesons composed of an $s \bar{s} $ pair.
We take a distinct value for the average number of $n_l$ in different mass ranges as in Eq.~\eqref{eq:dsigma}:
\begin{align}
    \overline{n}_l =
    \begin{cases}
        2 & (2m_l \le M < m_l + m_s)\,,\\
        \frac32 & (m_l + m_s \le M < 2m_s)\,,\\
        \frac43 & (2 m_s \leq M  < \infty)\,.
    \end{cases}
\end{align}
For $\bar{\sigma}$, we use the value, $\bar{\sigma} \simeq 30\MeV$, that is inferred from the fit to the lattice-QCD data for simplicity.
To incorporate the uncertainty, we vary $\bar{\sigma}$ by a factor two, i.e., we vary it between $\bar{\sigma} \simeq 15$--$60\MeV$.

\begin{figure}
    \centering
    \includegraphics[width=0.7\linewidth]{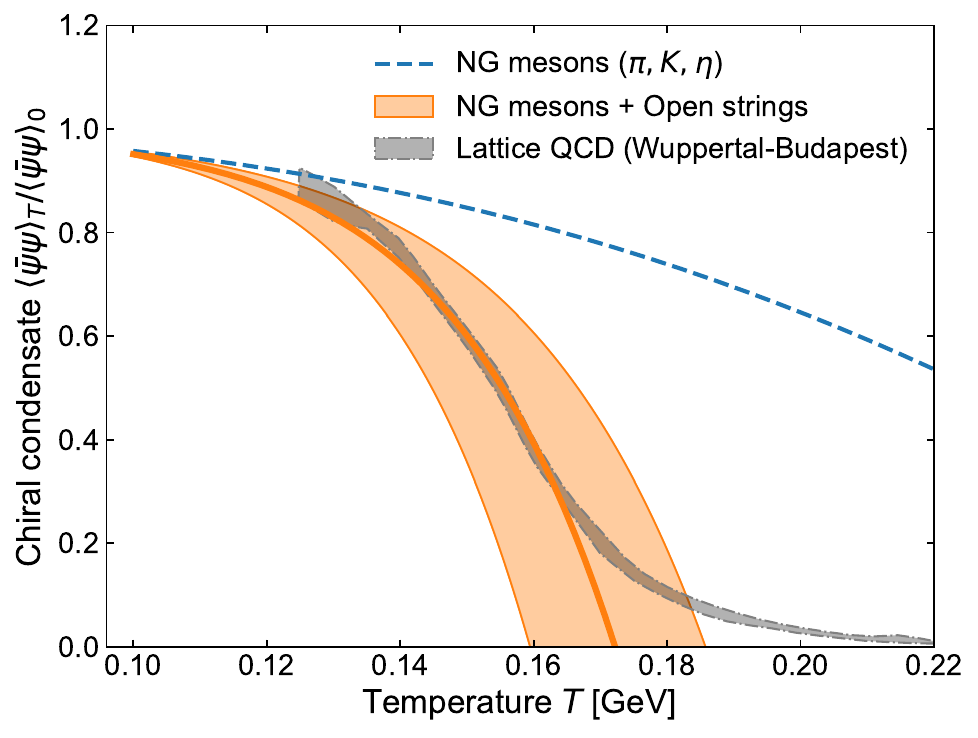}
    \caption{Ratio of the chiral condensate at finite temperature to that at zero temperature.
    The lattice data are taken from Ref.~\cite{Borsanyi:2010bp}.
    The averaged sigma term per quark, $\bar{\sigma} \simeq 30~\mathrm{MeV}$, is obtained from the best fit to the lattice-QCD data and it is varied within $15$–$60\MeV$ to account for uncertainty.
    The blue curve corresponds to the first term in the square bracket in Eq.~\eqref{eq:chiralT2}, and the orange curve includes the contribution from the second term in the square bracket in Eq.~\eqref{eq:chiralT2}.
    }
    \label{fig:chiral_condensate}
\end{figure}

The correction to $\langle \bar{\psi} \psi \rangle_T$ from an open string can be expressed as
\begin{equation}
\frac{\partial p_\text{open string}}{\partial m_q} 
= \int_{M_0}^\infty dM\, d_\sigma(M)\rho_{\text{open}}(M) \frac{\partial p(M)}{\partial m_q}\,,
\label{eq:chiralopenstring}
\end{equation}
where $d_\sigma(M)$, $\rho_{\mathrm{open}}(M)$, $\partial p(M) / \partial m_q$ are given by Eqs.~\eqref{eq:dsigma}, \eqref{eq:rhoopen}, and \eqref{eq:delP/delmq}, respectively.
By collecting the contributions from NG mesons and open strings, we can summarize the thermal correction to the chiral condensate as
\begin{align}
    \frac{\langle\bar{\psi}\psi\rangle_T}{\langle\bar{\psi}\psi\rangle_0} = 1 + \frac{m_q}{m_\pi^2f_\pi^2} \left[
    \sum_{i=\pi,K,\eta} g_i \frac{\partial p_i (M_i)}{\partial m_q}
    +\frac{\partial p_\text{open string}}{\partial m_q} \right]\,,
    \label{eq:chiralT2}
\end{align}
where $p_i$ and $g_i$ denote the pressure and the spin-flavor degeneracy, respectively, of the $i$-th hadron species with mass $M_i$, and we used the Gell-Mann--Oakes--Renner relation to replace $\langle\bar{\psi}\psi\rangle_0$ with $-m_q /(m_\pi^2 f_\pi^2)$ in the above expression.
In Fig.~\ref{fig:chiral_condensate}, we plot the resulting chiral condensate as a function of the temperature.
The sigma terms of the NG mesons are taken to be the same as those in Ref.~\cite{Biswas:2022vat}, and the lattice data are taken from Ref.~\cite{Borsanyi:2010bp}.

This is the demonstration for $\Nc = 3$ that the inclusion of the open string spectrum significantly alters the thermal behavior of the chiral condensate.
We note that this is also consistent with the previous observation that the chiral crossover can be well understood in terms of the HRG model, as argued in Ref.~\cite{Biswas:2022vat}.
We also note that the thermal correction in Eq.~\eqref{eq:chiralT2} can be $\mathcal{O}(\Nc^0)$, while the na\"{i}ve estimate in the chiral perturbation only predicts $\mathcal{O}(\Nc^{-1})$ correction.
This is because Eq.~\eqref{eq:chiralopenstring} shows the same $\Nc$-scaling as in the entropy density in Eq.~\eqref{eq:sstring}, thus inside the SQGB phase, it becomes $\partial p_{\text{open string}} / \partial m_q \sim \mathcal{O}(\Nc)$, which in total yields $\mathcal{O} (\Nc^0)$ correction to $\langle\bar{\psi}\psi\rangle_T/ \langle\bar{\psi}\psi\rangle_0$.

\section{Conclusions and Further Questions}

We have argued that the phases of QCD at finite temperature and density might include a new intermediate phase, which we named the Spaghetti of Quarks with Glueballs, the SQGB for short.  This is robust and generic to large $\Nc$, although many of our arguments are founded on phenomenological constraints that apply to the realistic $\Nc = 3$ case.  There is potentially interesting behavior relevant to phenomenology not only at the chiral restoration temperature, $\Tc$, but also at the deconfinement temperature for glueballs, i.e., $\Td \simeq 285\MeV$.
While the lattice-QCD data have been studied for various $\Nc$'s without quarks, they are also beneficial for us to gain an insight into the dynamical quark effects near the temperature region where the pure gluonic first-order deconfinement transition is washed out.  This is presumably the case for $\Nc$, not too far above $\Nc = 3$.

We have posed many questions with possible answers here, and another worthwhile question is how to characterize the region at finite baryon density properly.  Establishing the understanding there is complicated by the fact that the Hagedorn temperature for baryons may not be the same as that for mesons and glueballs~\cite{Broniowski:2000bj}.
This is because of the more complicated string structure of baryons compared to mesons and glueballs.  One more, and somehow related, question is how to characterize the region of transition between Quarkyonic Matter and the SQGB phase.  In the SQGB, chiral symmetry is likely to be restored, but in Quarkyonic Matter, chiral symmetry is broken by spatially inhomogeneous condensations.  It is an intriguing open question whether they are smoothly connected or separated by a sharp phase transition.

The SQGB phase is chirally symmetric but quarks are confined for large $\Nc$.  This seems contrary to the prejudice that chiral symmetry should be broken in confined systems.  The simplest heuristic argument for this is due to Banks and Casher~\cite{CASHER1979395,Banks:1979yr}.  For a quark-antiquark pair confined by vector-like interactions that do not induce spin flip, chirality must be flipped when a receding quark-antiquark pair changes its momentum due to confinement.  Strictly speaking, this is an argument for chiral $U(1)_\mathrm{A}$ symmetry breaking that is never restored exactly in QCD, and nevertheless, we might expect substantial breaking of $U(1)_\mathrm{A}$ chiral symmetry when the system is confined.  The question is;  what is the relation between chiral symmetry breaking and confinement~\cite{Alexandru:2019gdm}?  Might studying the SQGB phase for a wide range of $T$, $\muB$, and $\Nc$ resolve some of these issues?

Systematic studies of QCD at finite temperature for variable $\Nc$ for fixed number of flavors, and with variable masses can in principle clarify the issue of the relationship between the chiral transition and the Hagedorn temperature.  Such simulations could demonstrate or refute the existence of an intermediate phase of the type we discuss in this paper, particularly for large but finite $\Nc$.  At present, no evidence for such an intermediate phase is seen in such computations, but of course for the reasons stated above, such effects will be difficult to see due to the small effects of gluons/glueballs relative to quark/mesons, and because of lack of a divergence of the number of states in the glueball spectrum as one approaches $\Th$ from below~\cite{DeGrand:2021zjw}.

\acknowledgements
The authors thank Thomas~D.~Cohen and Leonid~Y.~Glozman for discussions and their efforts to organize ``Confinement and symmetry from vacuum to QCD phase diagram'' from Feb.~9 to Feb.~15, 2025, at Benasque Science Center, where this work was initiated.
Y.F.\ thanks Volker~Koch for discussions.
K.F.\ and L.M.\ thank Krzysztof~Redlich and Jean-Paul Blaizot for discussions.
We thank William A. Zajc for the careful reading of the manuscript and for the comments.
L.M.\ gratefully acknowledges many conversations with Gyozo Kovacs, Michal Marczenko, and Krzysztof Redlich at the university of Wroclaw.
Y.F.\ acknowledges support from the Japan Science and Technology Agency (JST) as part of Adopting Sustainable Partnerships for Innovative Research Ecosystem (ASPIRE) Grant No.\ JPMJAP2318, the National Science Foundation Grant No.\ PHY-2020275, and the Heising-Simons Foundation Grant 2017-228.
K.F.\ was supported by JSPS KAKENHI Grant No.\ 22H05118 and No.\ 23K22487.
Y.H.\ was supported by JSPS KAKENHI Grant No.\ 25K01002.
L.M.\ thanks partial support from the Institute for Nuclear Theory which is funded in part by the INT's U.S.\ Department of Energy grant No.\ DE-FG02-00ER41132.

\bibliography{sqgb}
\end{document}